# Single-photon detection timing jitter in a visible light photon counter

Burm Baek, Kyle S. McKay, Martin J. Stevens, Jungsang Kim, Henry H. Hogue, and Sae Woo Nam

*Abstract*— Visible light photon counters (VLPCs) offer many attractive features as photon detectors, such as high quantum efficiency and photon number resolution. We report measurements of the single-photon timing jitter in a VLPC, a critical performance factor in a time-correlated single-photon counting measurement, in a fiber-coupled closed-cycle cryocooler. The measured timing jitter is 240 ps full-width-at-half-maximum at a wavelength of 550 nm, with a dark count rate of $25 \times 10^3$ counts per second. The timing jitter increases modestly at longer wavelengths to 300 ps at 1000 nm, and increases substantially at lower bias voltages as the quantum efficiency is reduced.

## I. INTRODUCTION

HIGH-performance single-photon detectors (SPDs) that provide high quantum efficiency (QE), low dark-count rate and low timing jitter are crucial components in advancing quantum optics research. In addition to traditional photomultiplier tubes and single-photon avalanche diodes, several solid-state detectors have been demonstrated in recent years [1]-[4]. Among them, the visible light photon counter (VLPC) features a high QE (>80 %) and photon-number-resolving (PNR) capability with low multiplication noise at visible wavelengths [1], [5], [6]. Taking advantage of these properties, VLPCs have been used, for example, to demonstrate heralded photon-number-state generation [7].

Among SPDs with intrinsic PNR capability, only the transition edge sensor (TES) [8] outperforms the VLPC in both QE and PNR capability. However, a TES suffers from a larger timing jitter [~100 ns full-width-at-half-maximum (FWHM)] than most other SPDs [9], and requires a sophisticated cryogenic system to achieve the operating temperature of ~100 mK. The timing jitter of a detector directly affects the fidelity of a time-correlated single-photon counting measurement. Low timing jitter enables high-clock-rate quantum communication experiments and leads to improved signal-to-noise ratio (SNR) in gated photon-counting experiments by allowing short time-window sizes [10]. In this letter, we report a quantitative set of measurements on the single-photon detection timing jitter in a VLPC and discuss the device physics that explain these results.

## II. MEASUREMENT SYSTEM

We constructed a VLPC system equipped with a He gas closed-cycle cryocooler. The VLPC chip is mounted on a brass chip holder and wire-bonded directly to a coaxial connector. A brass fiber holder is used to align and hold the end of a standard single-mode optical fiber (SMF) directly above the VLPC, which has a circular active area of 1 mm in diameter. The gap between the fiber and the VLPC is ~1 mm, and the expected light spot size ($1/e^2$ width) varies from ~100 μm at 470 nm to ~200 μm at 1000 nm. The detector package, a thermometer, and a heater for temperature control are all mounted on a brass structure serving as the low-temperature stage. Temperature fluctuations are suppressed by using a thin insulator to weakly couple the detector stage and the cold head of the cryocooler. To reduce the blackbody radiation on the detector [1], a low-temperature (~4 K) shield is installed. A bias-T and a low-temperature amplifier are mounted on the intermediate cold stage (T ~ 40 K) for the voltage bias and readout. Another amplifier at room temperature further amplifies the VLPC pulses. In our measurement setup, output voltage pulses from the VLPC are ~1 ns FWHM, with a rising edge slope of ~250 mV/ns (inset in Fig. 1). We measured the pulse height distribution with excess noise factor $F = 1.03$ ($\equiv <M^2>/<M>^2$, $M$: multiplication gain), similar to that of previous reports [6], [11]. We also measured the dark count rate as a function of detection efficiency by varying bias voltage at a series of temperatures, and verified that these data all fall on a single characteristic curve of a nearly exponential shape in the dark count rate range from 20 counts per second (cps) to 25 kcps [4]. This implies that the background count rate from blackbody radiation or room light leakage is negligible in our system. Fig. 2 shows the measured system detection efficiency and dark count rate as a function of bias voltage at a temperature of 7 K and a wavelength of 633 nm. The peak system detection efficiency at 633 nm of 40% occurs at a bias voltage of 7.2 V

K. S. McKay was partially supported by NSF under CCF-0546068.

B. Baek, M. J. Stevens, and S. W. Nam are with the National Institute of Standards and Technology, Boulder, CO 80305 USA (e-mail: burm.baek@nist.gov).

K. S. McKay and J. Kim are with Fitzpatrick Institute for Photonics, Electrical and Computer Engineering Department, Duke University, Durham, NC 27708, USA.

H. H. Hogue is with DRS Sensors & Targeting Systems, Inc, 10600 Valley View St, Cypress, CA 90630, USA

and a temperature of 7 K. At higher bias voltages the dark counts continue to increase but the QE flattens out. The peak system detection efficiency is lower than the intrinsic QE of the VLPC (~95% [4]) but could be improved in the future with optimization of fiber coupling, reducing fiber connector losses, and anti-reflection coatings.

Fig. 1 shows a schematic of the experimental setup. We used a pulsed supercontinuum light source to measure timing jitter over a wide wavelength range. Optical pulses of ~100 fs duration from a Ti:Sapphire laser at 780 nm central wavelength and 82 MHz repetition rate pump a ~2 m long photonic crystal fiber to generate a supercontinuum of light pulses spanning from ~470 nm to ~1300 nm. After the fiber, a dichroic beamsplitter diverts most of the pump laser to a fast silicon photodiode, providing a reference clock synchronized with the laser pulse train that serves as the start for the timing electronics. The transmitted supercontinuum light is spectrally filtered to ~4 nm bandwidth by use of color filters and a grating monochromator, and then coupled into an SMF and guided to the VLPC. A time-interval analyzer (TIA) measures the VLPC timing jitter by recording a histogram of the time difference between clock pulses from the fast photodiode and output pulses from the VLPC. The TIA uses a constant-fraction discriminator, which reduces the effect of pulse-height distribution on the pulse-arrival timing. By replacing the VLPC with a superconducting nanowire single-photon detector, which has little wavelength-dependent jitter [12], we determined that the timing jitter of the measurement setup is less than 100 ps FWHM for all wavelengths. Electrical noise on the amplified VLPC pulses further degrades the total jitter by $\tau_{total} = (\tau_{noiseless}^2 + \tau_{noise}^2)^{1/2}$, where $\tau_{noiseless}$ is the jitter from an ideal noiseless measurement setup and $\tau_{noise}$ is the jitter that can be measured due the electrical noise by the TIA from a jitter-free signal. We estimate $\tau_{noise}$ ~100 ps FWHM.

### III. RESULTS AND DISCUSSION

We measured VLPC detection time distributions for wavelengths ranging from 470 nm to 1000 nm at a fixed temperature (7.0 K) and bias (7.2 V), adjusting the optical attenuation at each wavelength to keep the count rate ~50 kcps (average detected photon number per pulse $\mu = 6.3 \times 10^{-4}$). This count rate is well below the level at which performance degradation begins to occur due to saturation effects [4], [13]-[15]. Fig. 3(b) shows normalized, background-subtracted histograms of the VLPC at three wavelengths. The histograms approximately follow a Gaussian distribution at early time delays, followed by a long tail. Fig. 3(a) plots the extracted jitter FWHM over the full measurement range. The jitter increases slowly with increasing wavelength, from ~240 ps at 470 nm to ~300 ps at 1000 nm. The jitter increases rapidly with decreasing bias, from 210 ps to 480 ps as the bias is reduced from 7.6 V to 6.4 V at a wavelength of 633 nm and 7.0 K.

These characteristics can be understood by considering the operating principle of a VLPC. The VLPC consists of several silicon epitaxial layers grown on a degenerately doped Si substrate, as shown in Fig. 4(a), with total layer thickness of ~30 µm [16]. The gain and drift layers [distinguished only by the electric field profile, as shown in Fig. 4(b)] are highly doped with arsenic (donors) and slightly counterdoped with boron (acceptors). At operating temperature, the arsenic dopants in the gain/drift layer form a populated impurity band ~54 meV below the conduction band. At operating bias, an electric field in the VLPC is established, and field-assisted thermal ionization of donors leads to a stable "bias current" flowing through the device. The bias current sets up the region of constant electric field that defines the drift layer, crucial for achieving high QE [17]. The primary electron-hole pair generated by the absorption of a photon in the intrinsic layer is separated by the applied field, and the hole accelerates into the drift layer. The presence of the drift layer ensures that the hole has high probability of impact ionizing at least one neutral donor atom to provide an electron (referred to as a secondary electron) in the conduction band. The secondary electron drifts towards the front contact and initiates a local avalanche inside the gain layer, resulting in a pulse containing several tens of thousands of electrons. Long-wavelength visible and infrared photons may propagate beyond the intrinsic layer before they produce a primary ionization, in which case the primary electron will trigger the avalanche.

The electron drift velocity under voltage bias is shown schematically in Fig. 4(c). The mobility of the electrons in the drift and gain layers is limited by neutral-impurity scattering, and can be estimated using Erginsoy's formula (~1330 $cm^2V^{-1}s^{-1}$) [18]. Due to the low mobility and reduced field, the drift velocity of the secondary electron decreases rapidly in the drift layer. As the generation of a secondary electron occurs deeper in the drift layer, the delay between photon absorption and avalanche generation is prolonged, leading to an asymmetric shape and a long tail in the timing jitter distribution. At optimum bias condition, the impact ionization of secondary electrons occurs between the tail end of the gain layer and the top portion of the drift layer [marked by the arrow in Fig. 3(a)] [17], and the timing jitter, 240 ps at 470 nm, is attributed to this secondary-electron transit-time distribution convolved with our measurement system jitter (~100 ps).

The wavelength dependence of the VLPC jitter is a result of the distribution in transit times of the photogenerated holes and electrons before it triggers the avalanche. The hole drift velocity in the VLPC is ~1 $\times$ $10^7$ cm/s (for electric fields in a range of ~1-10 kV/cm) [19]. At a photon wavelength of 500 nm, the absorption coefficient (absorption length) of Si at the operating temperatures is about $5 \times 10^3$ $cm^{-1}$ (2 µm), whereas at 900 nm it is about $1 \times 10^2$ $cm^{-1}$ (100 µm) [17]. The timing jitter resulting from this hole transit-time spread is expected to be ~20 ps at 500 nm and ~130 ps at 900 nm. Near 900 nm, the photons are also absorbed in the gain and drift layers, where the primary electron will trigger the avalanche. The transit time before a primary electron generated deep in the drift layer can trigger an avalanche can be long (~1 ns) due to its low drift velocity, which increases the size of the tail in the distribution at longer wavelengths. When convolved with the timing jitter of ~240 ps associated with secondary-electron generation and its initiation of the avalanche, these mechanisms explain the wavelength dependence of the timing jitter.

The bias dependence of the VLPC jitter is associated with a lower electric field resulting in lower drift velocities and lower impact-ionization probability of the hole to generate the secondary electron [17]. Due to lower electron drift velocity in the drift region, the spread in the region of secondary-electron generation leads to a much broader timing distribution. The reduction of secondary-electron impact-ionization probability is also consistent with the reduction of the device QE as the bias voltage is reduced. A more quantitative model for the timing-jitter distribution is under development.

As the operating temperature is modified at a constant bias voltage, we find only a slight increase in the jitter with reduced temperature (~70 ps/K). As the temperature is reduced, the series resistance of the substrate increases resulting in a lower effective bias voltage. Decreasing the temperature is equivalent to reducing the bias voltage of the device and the temperature dependence of the jitter is consistent with the corresponding reduction in effective bias voltage [4].

### IV. Conclusion

Our study indicates that a VLPC system can achieve a timing jitter similar to that of common single-photon avalanche diodes (SPADs) at a reasonable dark count rate and detection efficiency [20] while providing PNR capability [5], [7] not offered by SPADs. A detailed understanding of the physical origins opens up the possibility for designing improved versions of VLPCs featuring better timing jitter characteristics.

### Acknowledgment

We thank Maryn "Dutch" Stapelbroek for his insights on VLPC operating principles, Jeff Van Lanen for technical assistance, and Scott Diddams for providing the photonic crystal fiber.

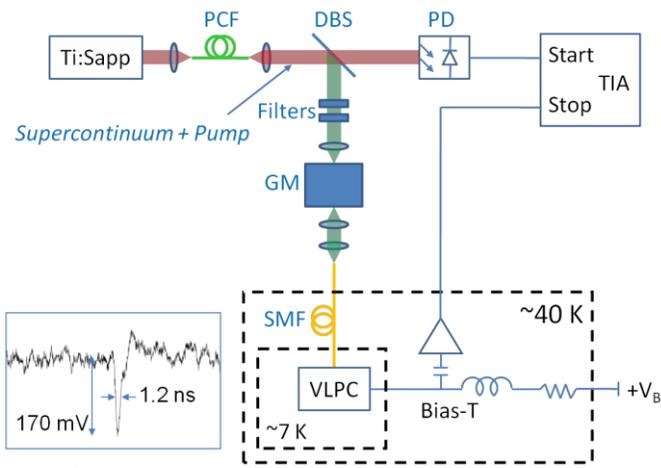

Fig. 1. Schematic of VLPC system and timing measurement. PCF: photonic crystal fiber, DBS: dichroic beamsplitter, Filters: color filters and neutral density filters, GM: grating monochromator, SMF: standard single-mode fiber, PD: fast photodiode, TIA: time-interval analyzer. Inset: VLPC pulse trace.

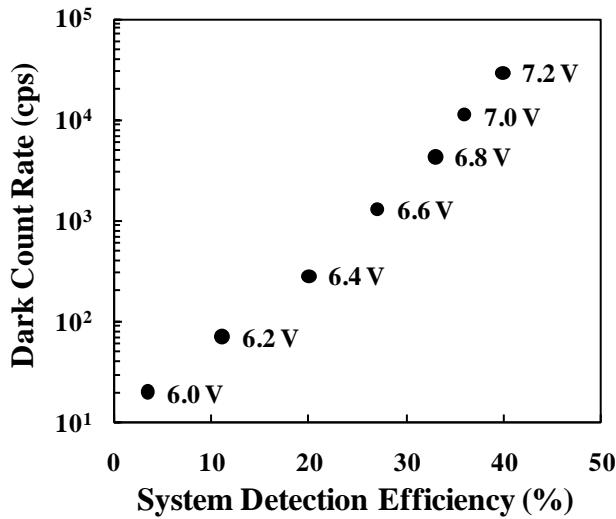

Fig. 2. The system detection efficiency and dark count rate as a function of bias voltage. The system detection efficiency was measured at a temperature 7 K and a wavelength of 633 nm.

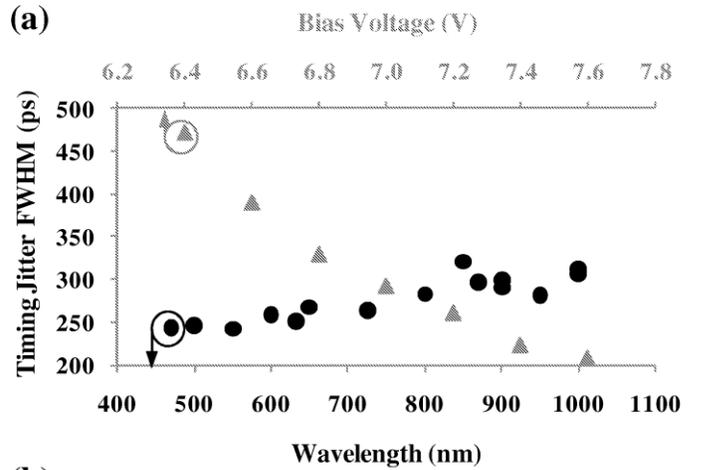

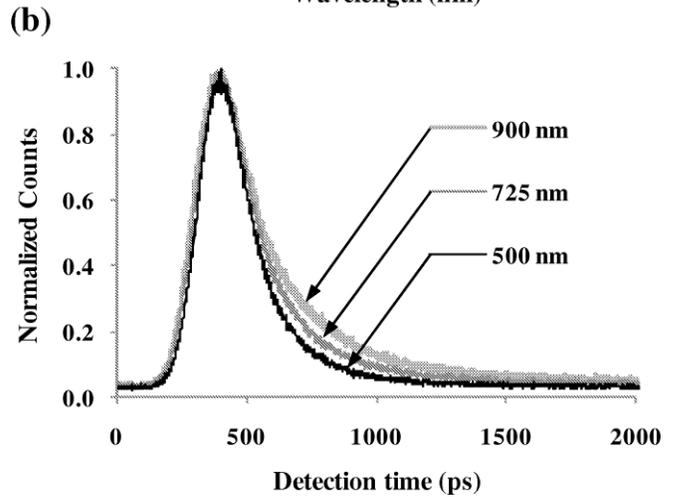

Fig. 3. Wavelength dependence (black circles) and bias voltage dependence (gray triangles) of the measured FWHM values of the timing jitter. The bias voltage dependence was measured at 633 nm and at a temperature of 7.0 K. The wavelength dependence was measured at a bias voltage of 7.2 V and at a temperature of 7.0 K. (b) Photon detection time histograms at different wavelengths.

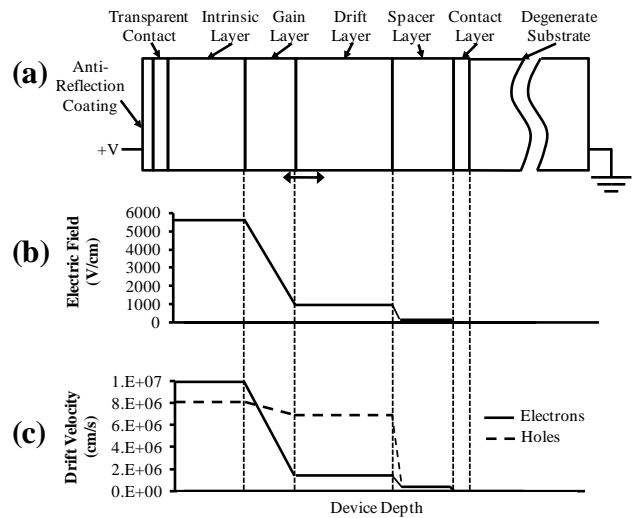

Fig. 4. Schematic of (a) VLPC structure, (b) calculated electric field distribution and (c) associated drift velocity of the electrons (solid line) and holes (dashed line) [19]. The total thickness of the shown epitaxial layers is ~30 μm [16] and the schematic is drawn roughly to scale.